\documentclass[twocolumn,showpacs,preprintnumbers,amsmath,amssymb]{revtex4-1}
\usepackage{graphicx}% Include figure files
\graphicspath{{figures//}}
\usepackage{color}% color text
\usepackage{hyperref} %%for creating hyperlink

\begin{document}

\title{Suppression of transverse instabilities of dark solitons and their dispersive shock waves}
\author{Andrea Armaroli and Stefano Trillo}
\email{stefano.trillo@unife.it}
\affiliation{Dipartimento di Ingegneria, Universit\`{a} di Ferrara, Via Saragat 1, 
44100 Ferrara, Italy}
\author{Andrea Fratalocchi}
\affiliation{
Dipartimento di Fisica, Universit\`{a} di Roma ``Sapienza'',  P. A. Moro 
2, 00185, Roma, Italy}

%claudio.conti@phys.uniroma1.it}
%\affiliation{
%$^{1}$ Research Center SMC INFM-CNR, Universit\`{a} di Roma ``La Sapienza'',  
%P. A. Moro 2, 00185, Roma, Italy\\
%$^{1}$Research center SOFT INFM-CNR Universit\`{a} di Roma ``La Sapienza'',  P. A. Moro 
%2, 00185, Roma, Italy\\
%$^{2}$Centro Studi e Ricerche ``Enrico Fermi'', Via Panisperna 89/A, 
%00184 Rome, Italy \\
%$^{3}$ Dipartimento di Ingegneria, Universit\`{a} di Ferrara, Via Saragat 1, 
%44100 Ferrara, Italy
%}

\date{30 Sept. 2009}  
\begin{abstract}
We investigate the impact of nonlocality, owing to diffusive behavior, on transverse instabilities of a dark stripe propagating in a defocusing cubic medium. 
The nonlocal response turns out to have a strongly stabilizing effect both in the case of a single soliton input and in the regime where dispersive shock waves develop (multi-soliton regime). 
Such conclusions are supported by the linear stability analysis and numerical simulation of the propagation.
\end{abstract}

%\pacs{42.65.Jx, 42.65.Tg, 47.40.Nm, 52.35.Tc}	% PACS, the Physics and Astronomy
% pacs aggiornati durante submission

\maketitle
\section{Introduction} 
Solitons confined in one dimension (1+1D) are generally unstable against periodic modulations in an extra dimension (1+2D) \cite{Kuznetsov86}. 
In particular, for the case of defocusing or repulsive Kerr-like (cubic) nonlinearities where the dynamics is governed by the equations of the nonlinear Schr\"odinger (NLS) type in two transverse dimensions, 1+1D solitons are of the dark type \cite{dark}. 
In this case, even if 1+1D dark solitons are stable against perturbations sharing 
the same dimensionality \cite{Barashenkov96}, the transverse instability in 1+2D determines breaking of dark stripes into a snake-like shape and subsequent decay into vortex-like structures \cite{Mamaev96,Tikh96,Anderson01,Feder00}, which is the reason why the mechanism is commonly known as {\em snake instability}. 
Snake-like breaking occurs also for bright solitons and other types of nonlinear response, and has been the object of recent interest \cite{Derossi97,Gorza04}.
Here we are interested in assessing the impact of a nonlocal Kerr response over the transverse snake instability of dark stripes.  
\newline \indent
A soliton has, by definition, an invariant profile due to perfect balance between nonlinearity and diffraction, and perturbations grow on top of that
over a characteristic length scale determined by the inverse growth rate of the instability.
Potentially, however, snake instabilities can also affect the propagation of a dark beam in the dynamical regime for which the nonlinearity overcompensates for diffraction. Particularly interesting, in this respect, is the strong nonlinear (or weakly dispersive) regime, where an initially smooth beam undergoes the formation of a dispersive shock wave (DSW), also known as undular bore (see, e.g. Refs. \cite{Gurevich74,Bronski94,Kodama99,Kamchatnov02,El05c,El07pra,Fratax08}).
In a DSW an infinite gradient (catastrophe), which develops owing to the initially dominant nonlinearity, is regularized by dispersion (diffraction), which determines the onset of  fast oscillations. These oscillations are basically trains of dark solitons \cite{Kamchatnov02,Fratax08} that fill a region which expands beyond the wave-breaking point, i.e. namely a shock fan, which is the true signature of the DSW. Such phenomenon, observed sporadically in the past \cite{Taylor70,Grischkowsky89}, have attracted recently a great deal of attention in the area of Bose-Einstein condensation (BEC) \cite{Dutton01,Hoefer06,Chang08} and nonlinear optics \cite{Wan07,DSWnonlocal,Jia07,Conti09}. 
In particular experiments in both fields demonstrate that DSW can occur from a dark input \cite{Dutton01,Conti09}. 
However, such results suggest that two different scenarios can occur, the first of which showing DSW to be unaffacted by transverse instabilities \cite{Conti09}, 
in contrast with an unstable situation where ultimate decay into vortices occurs \cite{Feder00,Dutton01,Chang08}. 
\newline \indent
In this respect, an important ingredient might be the nonlocality of the nonlinear response.  
In fact, since a nonlocal nonlinearity is known to induce suppression of modulational instabilities  \cite{DSWnonlocal,MInonlocal}  and transverse instabilities of bright solitons \cite{Lin08}, the same effect could be envisaged for dark inputs. 
Therefore it is important to investigate the onset of snake instabilities of dark soliton stripes and their DSW in nonlocal media. 
In this paper we pursue this goal by carrying out the analysis in the framework of a sufficiently general differential model
constituted by a NLS coupled to a diffusive equation  \cite{DSWnonlocal,Yakimenko05}.
\newline \indent
Concerning the link with previous results, we point out that suppression of transverse instabilities of a dark soliton stripe has been analyzed in the past, 
though in connection with other mechanisms such as saturation of nonlinearities, coupling to other modes or bright solitons (vector solitons), 
or incoherency of the propagating field \cite{Tikh96,suppression}.
Viceversa transverse instabilities of DSW are nearly unexplored. Investigation of instabilities of shock waves have been proposed theoretically in the framework of the Burger equation with small viscosity  \cite{Kuznetsov79}. Conversely, when the dynamics is ruled by weak dispersion (versus dissipation), 
only numerical studies are possible since transverse instabilities develop over a strongly dynamic, non-periodic evolution. In this context, our results show that an effective suppression of transverse instabilities of DSW generated by dark stripes is indeed possible in 1+2D.
\newline \indent
The paper is organized as follows. In Sec. \ref{model} we briefly discuss the model and its solitary waves of the dark type.
In Sec. \ref{lsa} we discuss the results of the linear stability analysis for a single soliton, while in Sec. \ref{vortex} we investigate 
the nonlinear stage of the instability showing vortex formation over long distances.
Finally in Sec. \ref{dsw}, we study numerically the onset of transverse instabilities for DSW.

\section{Nonlocal model and solitary solutions}
\label{model}

We investigate how nonlocality affects snake instabilities in the framework of the following dimensionless model,
which couples the nonlinear wave equation in the paraxial regime for the normalized  field $u(x,y,z)$ to a diffusive equation
which rules the refractive index change $n=n(x,y)$ 
\begin{eqnarray} \label{ODEsoliton}
\displaystyle i  \frac{\partial u}{\partial Z} +
\frac{1}{2} \nabla^2_{\perp} u - n u = 0, \label{nls1} \\
n - \sigma^2   \nabla^2_{\perp} n = |u|^2. \label{nls2}
%\end{array}
\end{eqnarray}
where $\sigma$ is a free external parameter which measures the degree of nonlocality. 
The local limit is obviously recovered for $\sigma=0$, for which Eqs.~(\ref{nls1}-\ref{nls2}) reduces to the standard integrable NLS equation.
The model has been shown to describe satisfactorily the behavior of thermal nonlinearities
in liquid solutions \cite{DSWnonlocal,Conti09}, as well as the dynamics in plasma \cite{Yakimenko05},
and has been used as a toy model for investigating different aspects of nonlocality \cite{Krow04,Kartashov07,Lin08}

\subsection{Dark solitons}

\begin{figure}[h]
\includegraphics[width=8cm]{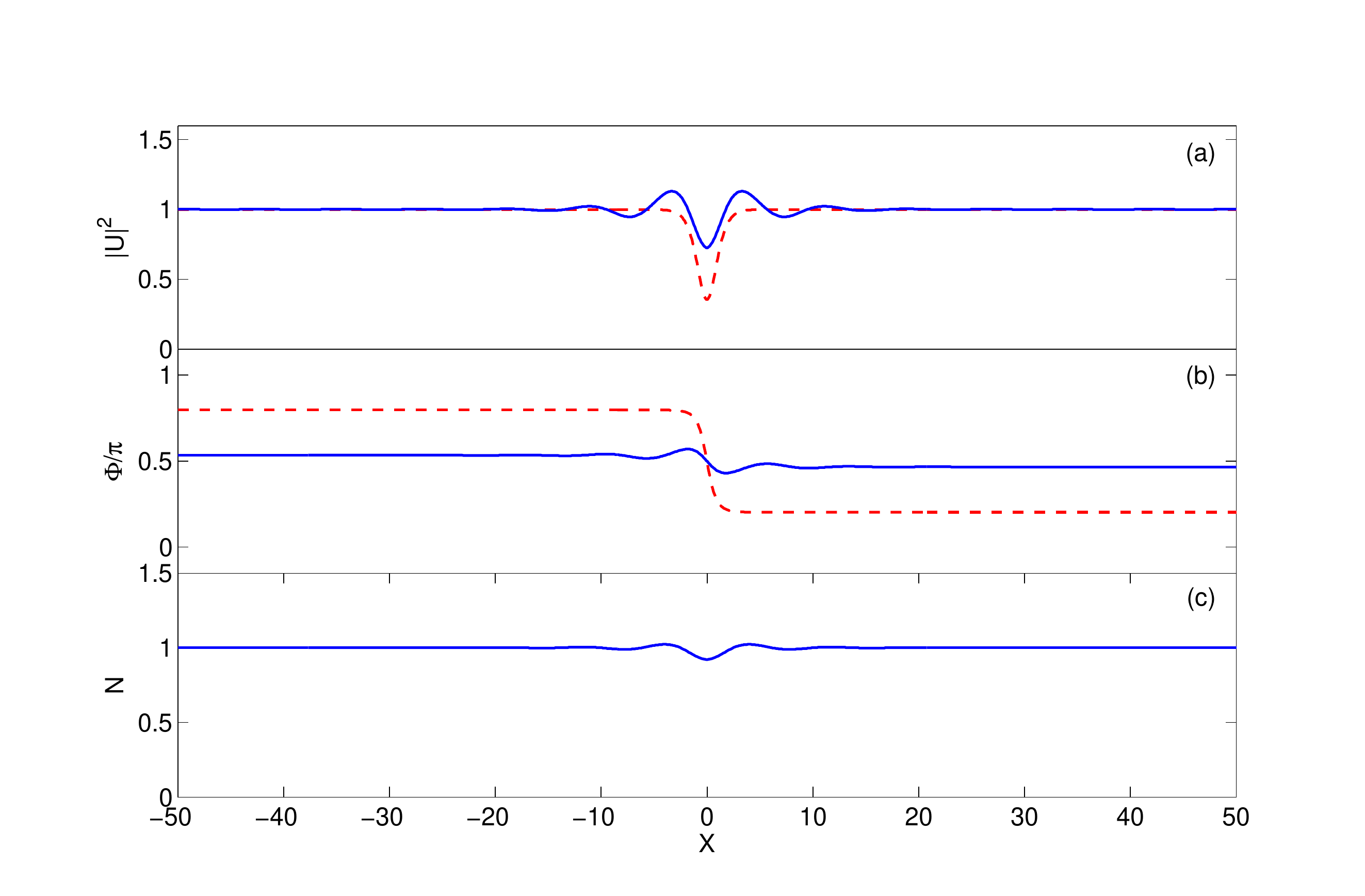}
\caption{(Color online) Example of dark soliton solution obtained numerically from Eqs.~(\ref{ode1}-\ref{ode2}) for $v=0.6$ and $\sigma^2=5$ (solid blue line), compared
with the corresponding solution of the local case $\sigma^2=0$ (red dotted line): (a) modulus of the field; (b) phase of the field; 
(c) refractive index change $N(X)$ across the soliton.}
\label{fig1} 
\end{figure}

Dark soliton solutions (strictly speaking solitary waves) of Eqs.~(\ref{nls1}-\ref{nls2}) have been investigated in Ref.~\cite{Kartashov07}.
Here we briefly recall the main results. 1+1D solutions are sought for in the form $u(x,z)={U}(X) \exp(i \beta Z)$, 
$n(x,Z)=N(X)$,  ${X} = x - v Z$ being the coordinate in the moving frame at velocity $v$. 
Here $U$ and $N$ are complex and real functions, respectively, which obey the coupled equations 
\begin{eqnarray}
\displaystyle \left( \frac{1}{2} \partial_{X}^2 -i v \partial_{X} - \beta - N \right)  {U} = 0, \label{ode1} \\
N - \sigma^2   \partial_{X}^2 N = |{U}|^2. \label{ode2}
\end{eqnarray}
%$(1/2 \partial_{\hat x}^2 -i v \partial_{\hat x} u_0 - \beta - n_0)  u = 0$ and $n_0 - \sigma^2   \partial_{\hat x}^2 n_0 = |u_0|^2$.
Without loss of generality we assume the plane-wave background of the soliton to have unit amplitude, 
i.e. $|{U}(X=\pm \infty)|=N(X=\pm \infty)=1$, which fixes the propagation constant to be $\beta=-1$.    
For fixed $\sigma$, with such boundary conditions, solitons are characterized by a single internal parameter, namely their velocity $v$.
The soliton family can be constructed by solving Eqs.~(\ref{ode1}-\ref{ode2}) numerically by means of the relaxation method, 
starting (for any fixed value of velocity $v$) from the explicit solution of the local NLS equation ($\sigma=0$) 
$U(X)=\sqrt{1-v^2} \tanh(\sqrt{1-v^2} X) + i v$, and iterating over the parameter $\sigma^2$. An example of solution obtained for $v=0.6$ and $\sigma^2=5$ is reported in Fig. \ref{fig1}, along with the corresponding local solution of the same velocity. As shown the soliton has a nonlinear phase profile with a net phase jump smaller than that of the local soliton. Both the modulus and the phase tend in a non-monotonic fashion to their asymptotic values characteristic of plane wave background. In particular for large $\sigma$ they both relax to the asymptotic values with damped oscillations. This is due to the fact that the eigenvalues of the linearization of Eqs.~(\ref{ODEsoliton}) around the fixed points (plane waves) are constituted by two complex conjugate pairs, with positive (unstable manifold) and negative (stable manifold) real part. For fixed $\sigma$, solitons exist below a critical maximal velocity $v_{max}$ ($0<|v|<v_{max}$), at which the real part of the eigenvalues vanish. We found this upper bound to the velocity to be described by the formula $v_{max}^2 =(4\sigma -1)/4\sigma^2$, which is valid for sufficiently large $\sigma$ ($\sigma^2>1/4$). Therefore nonlocal solitons exist for a more limited range of velocity compared with the local case. Moreover for any given velocity in the domain of existence the minimum intensity is larger than the corresponding local quantity. In other words local solitons are darker (their minimum intensity is closer to zero) than their nonlocal counterpart with the same velocity.
%darkness (minimum of $|{U}|^2$) turns out to be larger than the corresponding local quantity.

Finally, it is important to emphasize that nonlocal dark solitons turn out to be stable against perturbations in the same coordinate of confinement $X$, because the normalized momentum $P$ is always a decreasing function of velocity, i.e. $\partial P/\partial v <0$  \cite{Barashenkov96}.
%----------------------- FIGURE 1 ------------------------------
%-----------------------

\section{Linear stability analysis}
\label{lsa}

We proceed to analyze the stability of dark solitons against transverse perturbations by using the
following ansatz
\begin{eqnarray}
	u = \left[U(X) +  a(X,y,Z) \right] \exp(i \beta Z),\nonumber \\
	\label{eq:pert} \\
	n = N(X) + b(X,y,Z). \nonumber
\end{eqnarray}

%----------------------- FIGURE 2 ------------------------------
\begin{figure}[h]
\includegraphics[width=10cm]{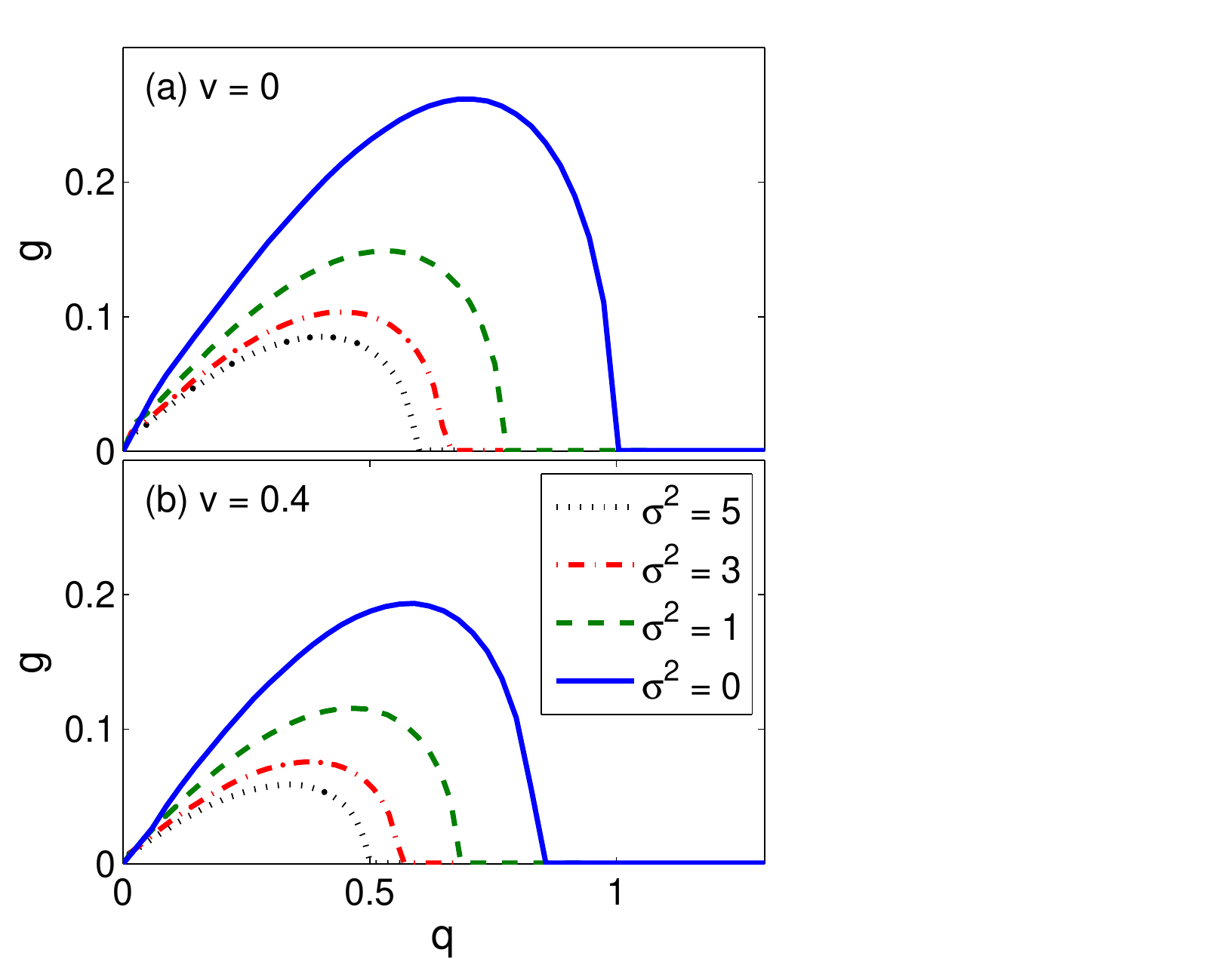}
\caption{(color online) Snake instability gain $g$ computed numerically vs. transverse wave-number $q$ for different values of nonlocality $\sigma^2$: (a) stationary case $v=0$; (b) $v=0.4$.}
\label{fig2} 
\end{figure}
%-----------------------

%----------------------- FIGURE 3 ------------------------------
\begin{figure}[h]
\centering
\includegraphics[width=8cm]{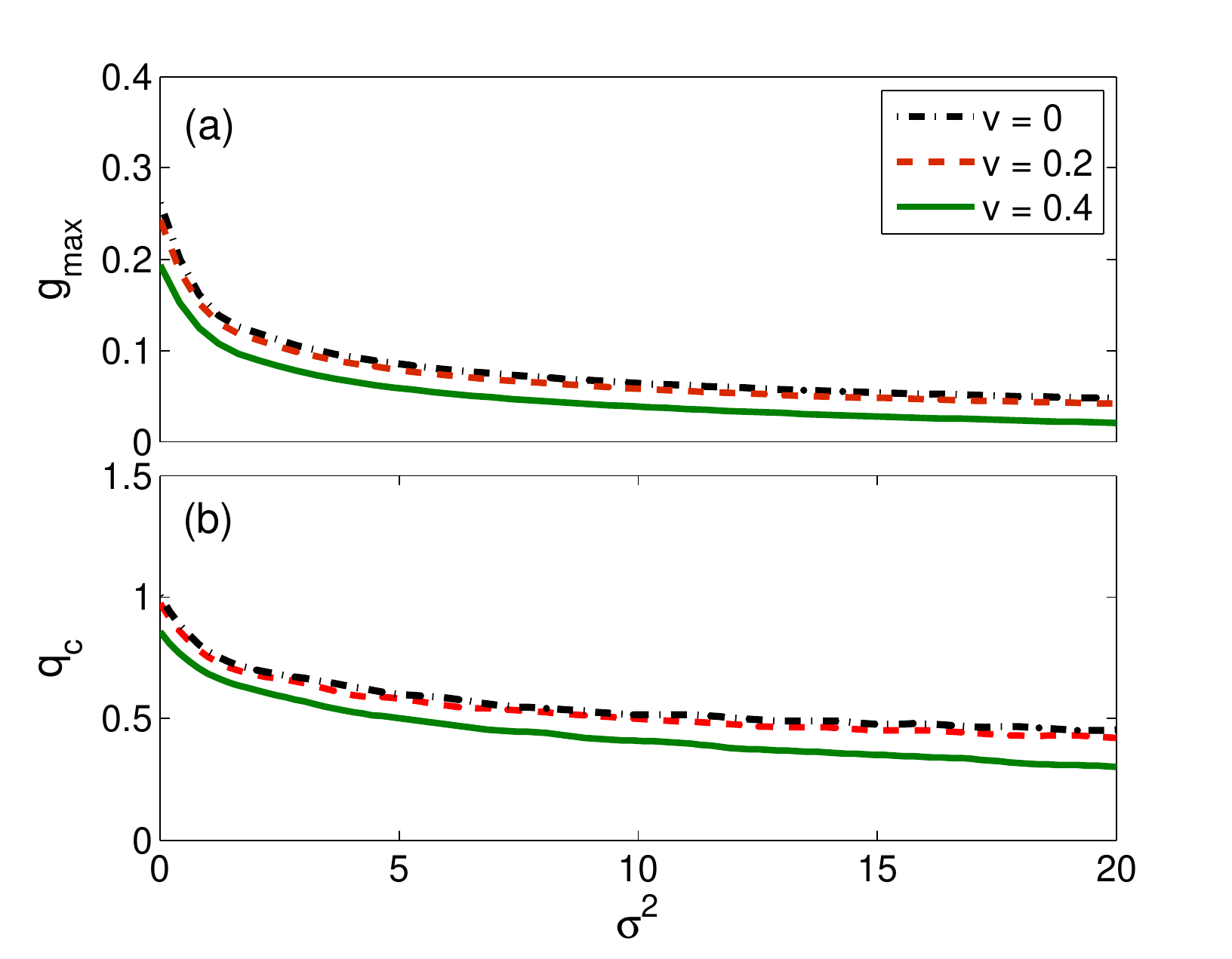}
\caption{(Color online) Overall reduction of the snake instability:
peak gain $g_{max}$ and cut-off (maximal) wave-number $q_c$ vs. $\sigma^2$ 
for three different velocities.}
\label{fig3} 
\end{figure}
%-----------------------

The perturbations $a,b$ ($a \ll U$, $b \ll N$) obey the linearized equations 
\begin{eqnarray}
%\begin{aligned}
	\left[ i \partial_z + \frac{1}{2}(\partial_{X}^2 + \partial_y^2) - i v \partial_{X} - \beta \right] a - \left( N a + U b \right)= 0,\nonumber \\
	\label{eq:MIlinearEVP} \\
	\left[ 1 - \sigma^2  (\partial_{X}^2 + \partial_y^2) \right] b =  U a^* + U^* a.\nonumber 
%\end{aligned}
\end{eqnarray}
%----------------------- FIGURE 4 ------------------------------
\begin{figure}
%\centering
%\hspace{-2cm}
\includegraphics[width=9cm]{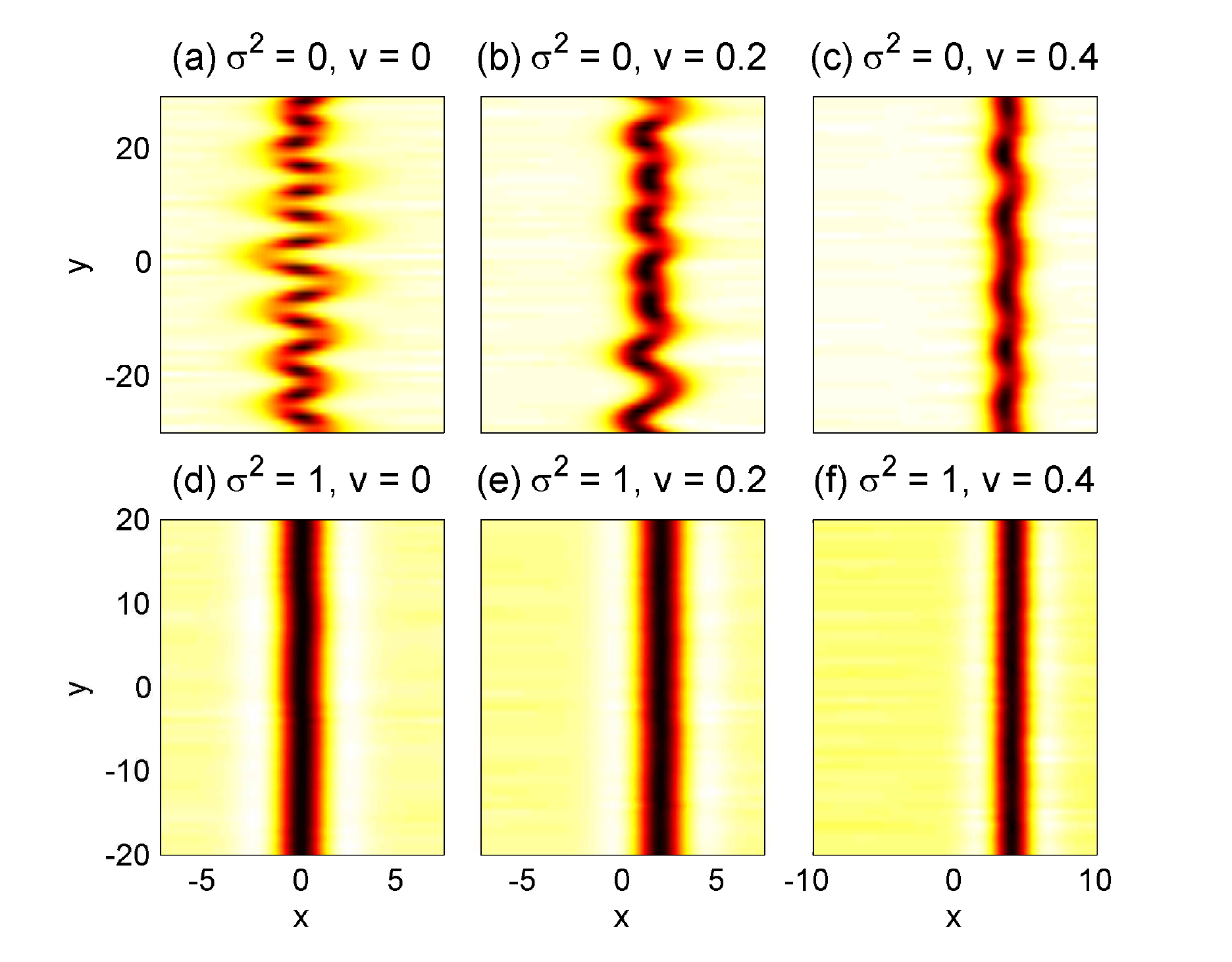} 
\hfill
%\hspace{2cm} \vspace{2cm}
\caption{(Color online) Results of numerical simulations based on Eqs.~(1-2):
intensity distribution over the output transverse plane at propagation distance
$Z=10$ for different values of $\sigma$ and $v$.
Top row, local case $\sigma=0$: (a) $v=0$; (b) $v=0.2$; (c) $v=0.4$.
Bottom row, nonlocal case $\sigma=1$: (d) $v=0$; (e) $v=0.2$; (f) $v=0.4$.}
\label{fig4} 
\end{figure}
%-----------------------
Then we write the perturbations in terms of sideband pairs with transverse wave-number $\pm q$ as 
\begin{eqnarray}
%\begin{aligned}
	a&=&a_+(X) \exp(i qy + i\lambda Z) + a_-(X) \exp(-i qy - i\lambda^* Z), \nonumber \\
	\label{eq:sidebands} \\
	b&=&b_+(X) \exp(i qy) + b_-(X) \exp(-i qy),\nonumber 
%\end{aligned}
\end{eqnarray}
where the additional constraint $b_-=b_+^*$ must be imposed for the index perturbation to be real.
By inserting Eqs.~(\ref{eq:sidebands}) in Eqs.~(\ref{eq:MIlinearEVP}) and defining the diffusion linear operator
$\mathcal{L}_d \equiv 1 + \sigma^2 q^2 -\sigma^2 \partial_{X}^2$ and its inverse $\mathcal{L}_d^{-1}$, 
we obtain from the second of  Eqs.~(\ref{eq:MIlinearEVP})
\begin{equation}
	b_+ = \mathcal{L}_d^{-1} \left( U a^* + U^* a \right)
\end{equation}
By substituting this expression into the first of Eqs.~\eqref{eq:MIlinearEVP}, and defining the linear differential operator
%\begin{equation}
$$\mathcal{L}_0 \equiv \frac{1}{2} \partial_{X}^2  - \beta - \frac{q^2}{2}  - N(X),$$
%\end{equation} 
as well as the unknown array (sideband amplitudes) ${\bf a} = \left[ a_+ \; a_-^* \right]^T$, 
%we search for solutions with dependence ${\bf a} \sim \exp(i \lambda Z)$ 
we obtain the following canonical eigenvalue problem
\begin{eqnarray} \label{eq:EVP}
A {\bf a} = \lambda {\bf a},
\end{eqnarray}
where the matrix $A$ reads as
\begin{eqnarray*} %\label{eq:linop}
	A=
	\begin{bmatrix}
		\mathcal{L}_0- i v \partial_{X} - {U} \mathcal{L}_d^{-1} {U}^* & -{U} \mathcal{L}_d^{-1} {U}\\
		{U}^* \mathcal{L}_d^{-1} {U}^*  & -\mathcal{L}_0 - i v \partial_{X} + {U}^* \mathcal{L}_d^{-1} {U}
	\end{bmatrix}.
\end{eqnarray*}
We solve the eigenvalue problem (\ref{eq:EVP}) numerically by introducing a suitable discretization of the operators
$\mathcal{L}_0$, $\mathcal{L}_d$ along the X axis. Instability occurs for negative imaginary eigenvalues, entailing a perturbation
that grows exponentially with rate $g = -\operatorname{Im}(\lambda)$. The eigenfunctions (not shown) that correspond to unstable eigenvalues have even symmetry, 
which associated with the odd symmetry of the soliton, leads to a snake type of instability.
The dependence of the snake instability gain $g$ on the wave-number $q$ is displayed in Fig.~\ref{fig2} 
for different values of $\sigma$ both in the stationary ($v=0$, still black solitons) and non-stationary case ($v=0.4$, moving solitons). 
As shown, when nonlocality becomes progressively more effective (i.e., when $\sigma$ is increased), 
a substantial suppression of the instability is observed. 
In fact, both the peak gain $g_{max}=\max[g(q)]$ and the cut-off wave-number $q_c$ (below which $g \neq 0$) decrease as $\sigma$ grows larger.
The dependence of peak gain $g_{max}$ and cut-off wave-number $q_c$ on the nonlocal parameter $\sigma$,
for three different representative velocities $v=0,0.2,0.4$, is summarized in Fig.~\ref{fig3}.

In order to verify that the snake instability is effectively suppressed during propagation of a dark stripe, 
we resort to numerical integration of Eqs.~(1)-(2) by split-step method. 
We use initial data $u_0=u(x,y,z=0)=U(x) + \xi(x,y)$,
$\xi(x,y)$ denoting henceforth a noise term ($|\xi(x,y)| \ll |U(x)|$) with random (normally distributed) modulus and phase. 
%with variance $\sigma_n^2=<\xi(x,y)^2>=$.
The results of the simulations, comparing the local ($\sigma=0$) and the nonlocal ($\sigma=1$) case, for three different values of velocity
are reported in Fig.~\ref{fig4}. 
In the local case, one observes the onset of snake instabilities for any value of velocity (the instability is weaker for larger velocities), 
as clearly shown in Figs. \ref{fig4}(a-c).
However, in the nonlocal case, as shown in Fig.~\ref{fig4}(d-f), the instability is effectively suppressed and the stripe symmetry of the input is preserved. 

It is worth emphasizing that we find a similar stable behavior (results not shown) also for bound states formed by two soliton stripes \cite{Kartashov07}.
However, in this case, it is improper to talk about the stabilizing role of nonlocality, since these states do not survive in the limit of local nonlinearities, being peculiar to the nonlocal response.

\section{Vortex solitons and their generation}
\label{vortex}
%----------------------- FIGURE vortex 1 ------------------------------
\begin{figure}[h]
\hspace{-0.5cm}
\includegraphics[width=8cm]{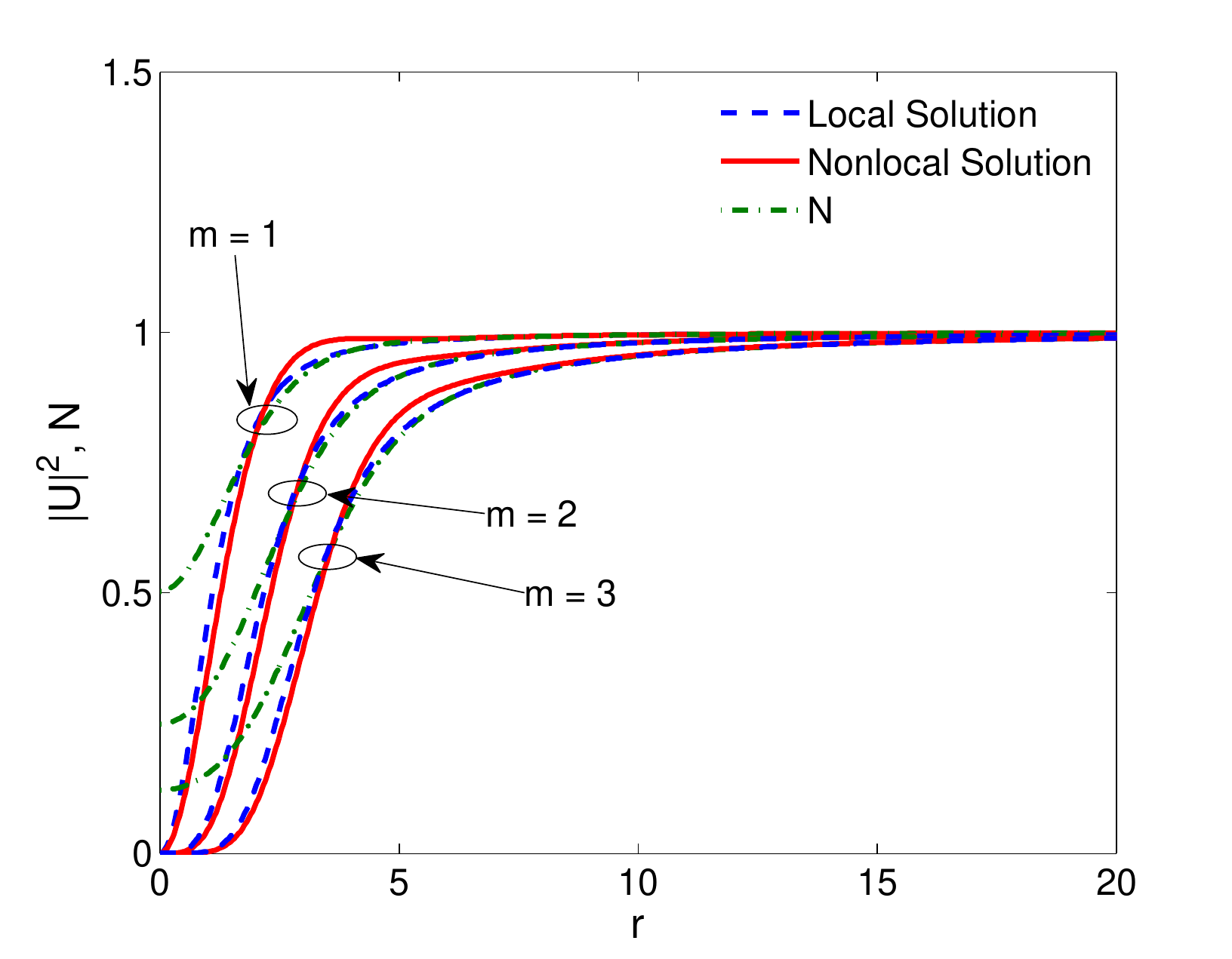} 
\caption{(Color online) Radial intensity profiles (red  solid line) of nonlocal vortex solutions  and relative index profile (green dot-dashed line)
for different topological charges $m=1,2,3$, 
and $\sigma=1$. For comparison we report also the local intensity profile (dashed blue line).
}
\label{vortex1} 
\end{figure}
%-----------------------

%----------------------- FIGURE vortex 2 ------------------------------
\begin{figure}[h]
\hspace{-0.5cm}
\includegraphics[width=12cm]{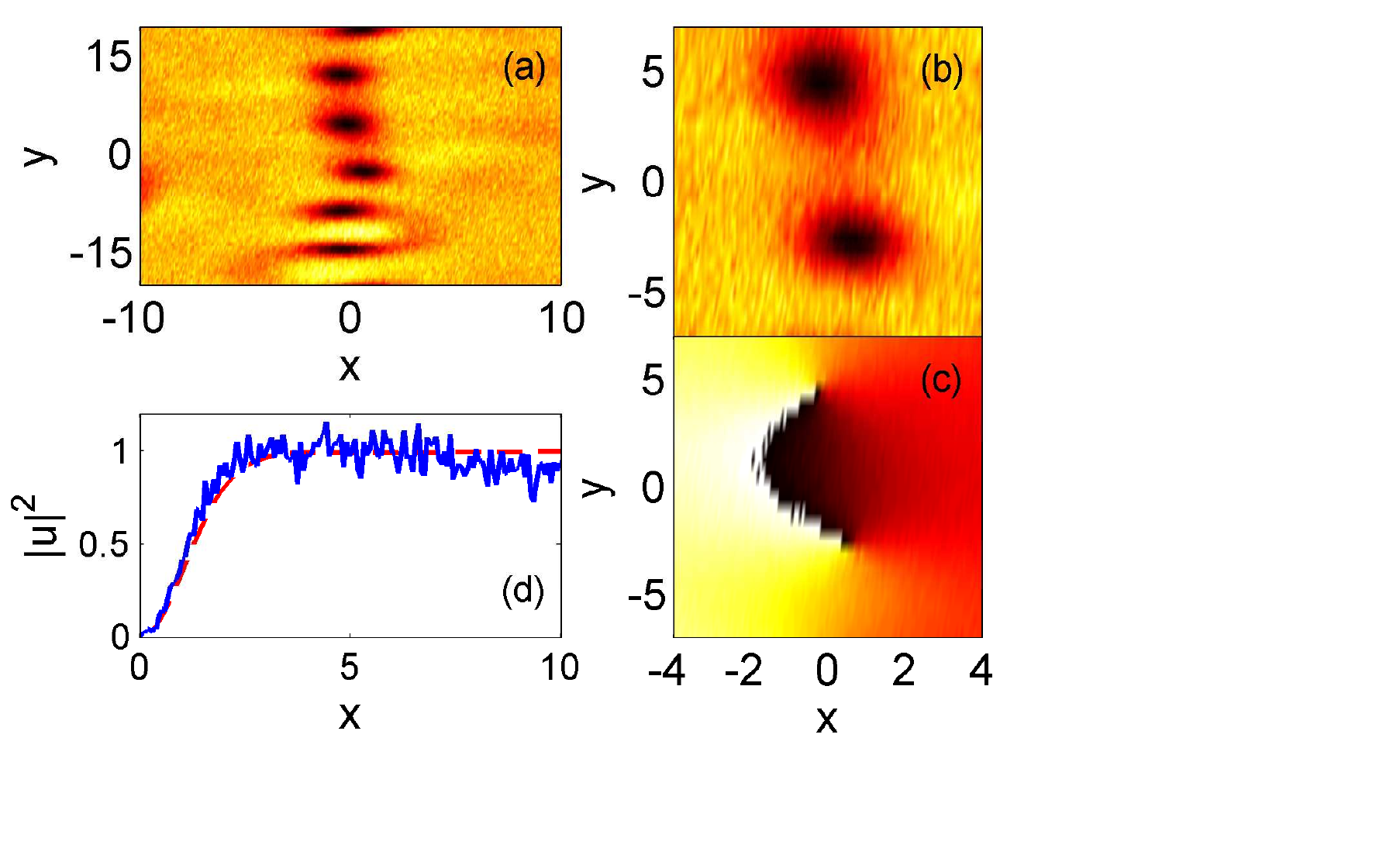} 
\caption{(Color online) Results of numerical simulations based on Eqs.~(1-2) of long term evolution of an input stationary dark soliton stripe.
(a) intensity distribution over the output transverse plane at $Z=80$.
(b-c) insets showing a portion of the transverse plain including two adjacent vortex solitons: 
(b) intensity distribution; (c) phase distribution, showing opposite helicity. 
(d) noisy intensity profile along $x$ (blue solid line) of the upper vortex in (b) as seen at fixed $y=5$, 
compared with the theoretical profile (red dashed line) of the vortex mode as obtained from Eqs. (10-11).
Here $\sigma=1$. 
}
\label{vortex2} 
\end{figure}
%-----------------------
Having shown that nonlocal dark stripes remain stable over characteristic distances where their local counterparts break down, 
one might wonder about propagation over much longer distances (even though these can be hardly reached in practical experiments).
More specifically it is natural to ask whether the weak residual instability determines a decay scenario qualitatively similar to the
local one, where an unstable dark stripe leads to the generation of vortex solitons \cite {Mamaev96,Tikh96}. 
In order to investigate this point, we have first searched for vortex solutions of the nonlocal model, and then explored whether 
they are generated through the long-range instability of the dark stripe.

Stationary vortex solitons are characterized by a field $u(r,\varphi,z)=U(r) \exp(i m \varphi + i \beta z)$, where  the topological charge $m$ fixes the azimuthal dependence (helicity of the phase) and $r=\sqrt{x^2 + y^2}$ is the radial variable. The associated index $n(r,\varphi)=N(r)$ has only radial dependence since it depends on intensity. 
$U$ and $N$ obey the following equations
\begin{eqnarray}
\displaystyle \left[ \frac{1}{2} \left( \partial_{r}^2 + \frac{1}{r} \partial_{r} - m^2 \right)  - \beta \right] \,U - N \,U = 0, \label{v1} \\
N - \sigma^2   \left( \partial_{r}^2 + \frac{1}{r} \partial_{r}  \right) N = |{U}|^2, \label{v2}
\end{eqnarray}
which can be solved numerically by the relaxation method.
%Eqs.  (\ref{v1}- \ref{v2})
An example of the radial profile of such solutions, obtained for a fixed value of nonlocality $\sigma=1$ and different helicities $m=1,2,3$,
is displayed in Fig. \ref{vortex1}. As shown, the nonlocal profiles are qualitatively similar to the corresponding local vortex solitons,
except for the fact the index change does not vanish in $r=0$ (where the field instead vanishes) because of the averaging due to the nonlocality.

In order to show that these vortex solitons are stable modes of the system and appear spontaneously due to the residual instability of the dark stripe,
we have extended the simulations of Fig. \ref{fig4} to much longer distances. A typical result, obtained for $\sigma=1$, is reported in Fig. \ref{vortex2}.
As shown in Fig. \ref{vortex2}(a), at sufficiently long distances, the soliton breaks down into an irregular sequence of dark spots. 
A closer look to two adjacent spots [see insets in Figs. \ref{vortex2}(b-c)] reveals that they are indeed vortex pairs with opposite charge (helicity). 
The radial profile of the spontaneously generated vortex modes, though affected by the noise, are closely reminiscent of the solutions
of Eqs. (10-11), as clearly shown in Fig. \ref{vortex2}(d). 

The results shown in Fig. \ref{vortex2} allows us to conclude that the residual instability leads to a scenario dominated by spontaneous formation
of vortex pairs, similarly to the local case. In this case, however, much longer distances are needed to observe the phenomenon because of 
the substantial weakening of the instability due to the nonlocality.

\section{Instabilities of dispersive shock waves}
\label{dsw}

%----------------------- FIGURE 5 ------------------------------
\begin{figure}[h]
\hspace{-0.5cm}
\includegraphics[width=4.3cm]{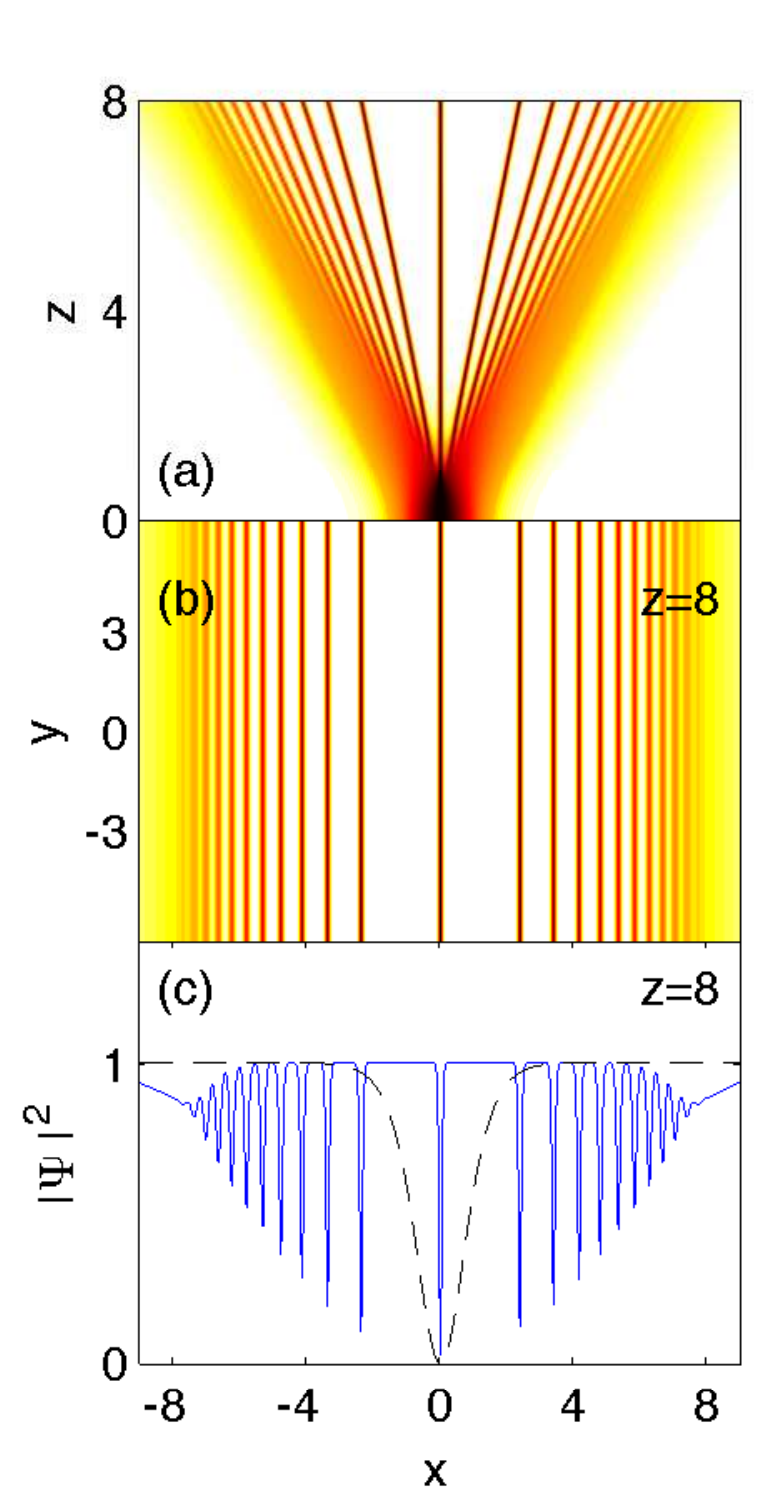} 
\includegraphics[width=4.3cm]{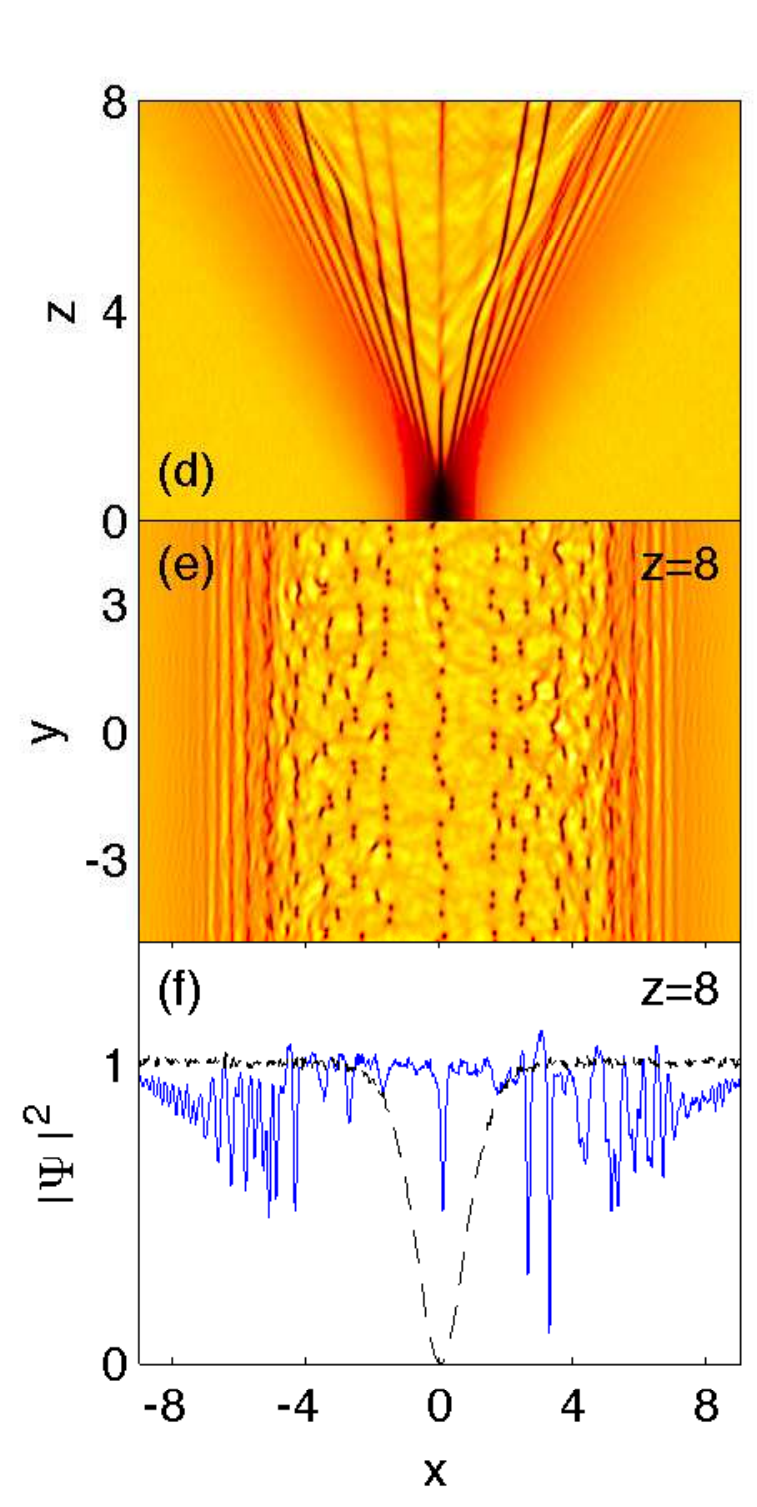}
%\includegraphics[width=4.3cm]{figpra/loc2D_v0_unpert} 
%\includegraphics[width=4.3cm]{figpra/loc2D_v0_noise}
% from data loc2D_v0_eps0p05_n0p005cmplx ix=3*ny/8 and xshift=-15 (left soliton)
% 256x2048 on transverse plane
\caption{(Color online) DSW ruled by the local 1+2D NLS equation, 
contrasting the ideal noise-less case (a-b-c) 
with the evolution in the presence of noise (d-e-f).
(a-d) Shock fan at $y=0$; (b-e) Output transverse plane; 
(c-f) Output intensity profile (solid line) at $y=0$ superimposed to the input (dashed line).  Here $\varepsilon=0.05$, 
and the soliton stripe input is $U(x) = \tanh(x)$.}
\label{DSWloc} 
\end{figure}
%-----------------------

%----------------------- FIGURE 6 ------------------------------
\begin{figure}[h]
\includegraphics[width=15cm]{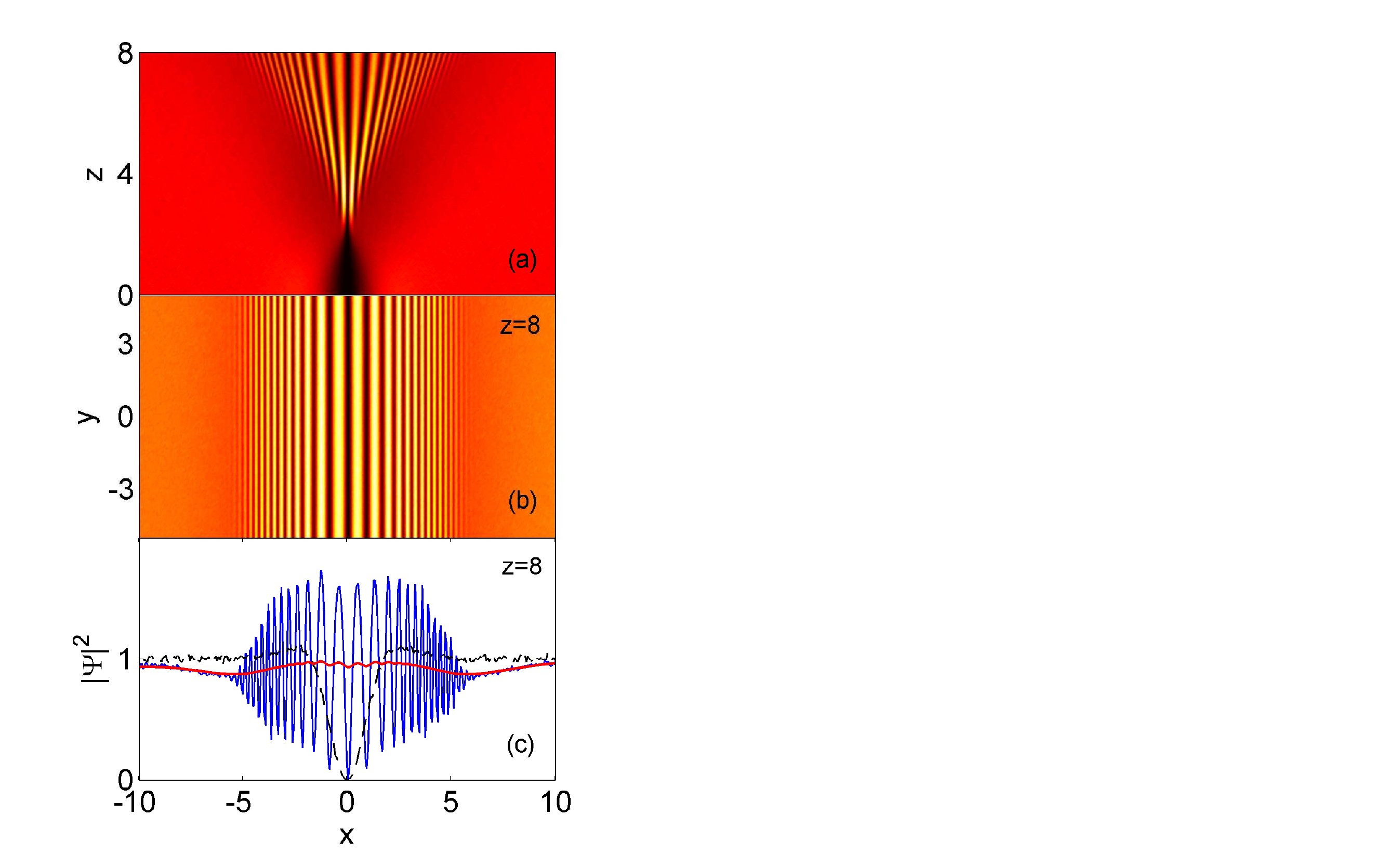}
% from data by Andrea A.
%\includegraphics[width=5cm]{figpra/nloc_v0_ky0p5}
%\includegraphics[width=4.4cm]{nloc_v0_andrea}
% from nonlocalshockTS_v0_soliton1.fig
\caption{(Color online) As in Fig. 5, showing stabilization of a DSW against transverse instability, in the nonlocal regime.
Results from numerical integration of Eqs.~(\ref{nls3})-(\ref{nls4}) with $\sigma=1$ and $\varepsilon=0.05$. (a) Shock fan at $y=0$; (b) Output ($z=8$) transverse plane; 
(c) Output intensity profile (thin solid blue line) and refractive index change $n(x)$ (thick solid red line) at $y=0$ superimposed to the input (dashed line).}
\label{DSWnloc} 
\end{figure}
%-----------------------

When the intensity becomes much higher than that needed to support a soliton, nonlinearity overcompensates for diffraction, 
and eventually leads a smooth input to develop an infinite gradient (catastrophe) which is regularized by diffraction, forming a DSW.
In the case of a hyperbolic tangent input the catastrophe occurs in the beam center, and post-shock oscillations emerge,
in the dynamics governed by the local 1+1D NLS, in the form of a soliton train, with pairs of dark solitons 
emerging at progressively larger angles (transverse velocity) around a central immobile soliton. 
A qualitatively similar scenario takes place in the nonlocal case \cite{Conti09}.
Here we investigate the stability of such DSW structures against transverse perturbations in 1+2D. 
The highly dynamical non-periodic character of the 1+1D evolution does not allow to extend the linear stability analysis
carried out for single solitons, and hence we resort to numerical simulations. 
In particular, we restrict ourselves to input conditions $u_0=u(x,y,z=0)=\rho \left[ U(x) + \xi(x,y) \right]$, where $U(x)$ is
the 1+1D (either local or nonlocal) soliton, and $\rho \gg 1$ is the factor which measures ovecompensation (i.e., deviation from soliton case).
%$\xi(x,y)$ is a normally distributed white noise term. 
However, in order to make evident that, here, we operate in the weakly dispersive limit, we introduce the scaling commonly 
adopted in the theory of DSW \cite{Conti09}, casting Eqs.~(1)-(2) in the semiclassical form
\begin{eqnarray}
%\begin{array}{l}
\displaystyle i \varepsilon  \frac{\partial \psi}{\partial z} +
\frac{\varepsilon^2}{2} \nabla^2_{\perp} \psi - \hat n \psi=0, \label{nls3} \\
\hat n - \sigma^2   \nabla^2_{\perp} \hat n = |\psi|^2, \label{nls4}
%\end{array}
\end{eqnarray}
which is obtained from Eqs.~(1)-(2) with the positions $z=Z/\varepsilon$, $\hat n = n \varepsilon^2$,
$\psi=\varepsilon u$, where $\varepsilon=1/\rho \ll 1$. Such scaling allows us to keep the input fixed to
$u_0=U(x) + \xi(x,y)$ and increase the nonlinearity by decreasing $\varepsilon$. 
In the following we present specific examples obtained for $\varepsilon=0.05$, 
using split-step method with a typical transverse grid of $2048 \times 256$ points  ($x \times y$) over the range of interest.
Note that, in order to employ periodic boundary conditions along the confinement coordinate $x$, the calculations are carried out over a window much larger 
than that shown in the figures below, containing two solitons with opposite phase and sufficiently far from each other, so to ensure negligible interaction.

The formation of a DSW occurring in the local limit $\sigma=0$ from a black soliton initial shape ($U(x)=\tanh(x)$, corresponding to $v=0$) 
is illustrated in Fig.~\ref{DSWloc}. For reference we display in Fig.~\ref{DSWloc}(a-c) the ideal DSW obtained by integrating Eqs.~(\ref{nls3})-(\ref{nls4}) in the absence of noise. We recall that  the occurrence of wave-breaking can be described in terms of hydrodynamical variables $h=|\psi|^2$ (water height or density of an equivalent fluid) and $V=\varphi_x$ (velocity of the water or the equivalent fluid),  
which obey the hyperbolic equations (see, e.g. Refs.~\cite{Hoefer06,Wan07,DSWnonlocal,Conti09})
\begin{eqnarray}
h_{z}  + \left(h u \right)_{x} = &0&, \label{wkb1} \\ 
u_{z} + u u_{x} + h_{x} = &0&, \label{wkb2}
\end{eqnarray}
obtained at leading order by applying the transformation $\psi=\sqrt{h} \exp(i \varphi / \varepsilon)$ to Eqs.~(\ref{nls3})-(\ref{nls4}) in the local limit. 
In the case shown the gradient catastrophe occurs at the point of vanishing intensity $x=0$. 
The catastrophe becomes manifest at finite distance  ($z=z_s \simeq 0.75$) as a vertical front in the velocity $V$ accompanied by a cusp in the intensity (density)  \cite{Conti09}.
Beyond such wave-breaking point, diffraction along the $x-$coordinate regularizes the infinite gradient, 
and a DSW becomes apparent as a fan [see Fig.~\ref{DSWloc}(a)], which is progressively filled with transverse oscillations taking the form of narrow gray soliton filaments.
Asymptotically  ($z \sim 1/\varepsilon$), the number of emerging solitons is $2 \rho -1$ for $\rho = \varepsilon^{-1}$ integer, 
and the velocities of symmetric pairs tend to accumulate towards linear velocity $|dx/dz|=1$ \cite{Fratax08}.
In the absence of noise the solitons in the fan emerge onto the transverse plane as stripes, as shown in Fig.~\ref{DSWloc}(b).
However, such scenario changes radically when noise is introduced, as illustrated by Fig.~ \ref{DSWloc}(d)-(e)-(f).
Transverse instabilities  rowing from noise cause indeed the soliton stripes to break [see Fig.~\ref{DSWloc}(e)]. 
Solitons with larger darkness (i.e., dip of intensity closer to zero) break down at shorter distances, consistently
with the fact that the gain decreases for larger velocity, as can be inferred by comparing the gain curves in Fig. 2(a) and 2(b).
The onset of the instability makes the shock fan, once seen on the plane $y=0$, irregularly filled with erratic filaments which change velocity in unpredictable manner, as clear from Fig.~\ref{DSWloc}(d). The instability not only results into a strongly irregular train along $x$ [compare Fig.~ \ref{DSWloc}(c) and Fig.~ \ref{DSWloc}(f)], 
but also into the fact that such train looks different when seen at different elevations (i.e., planes $y=constant$).
%----------------------- FIGURE 7 ------------------------------
 \begin{figure}[h]
\centering
%\hspace{-1cm}
\includegraphics[width=8cm]{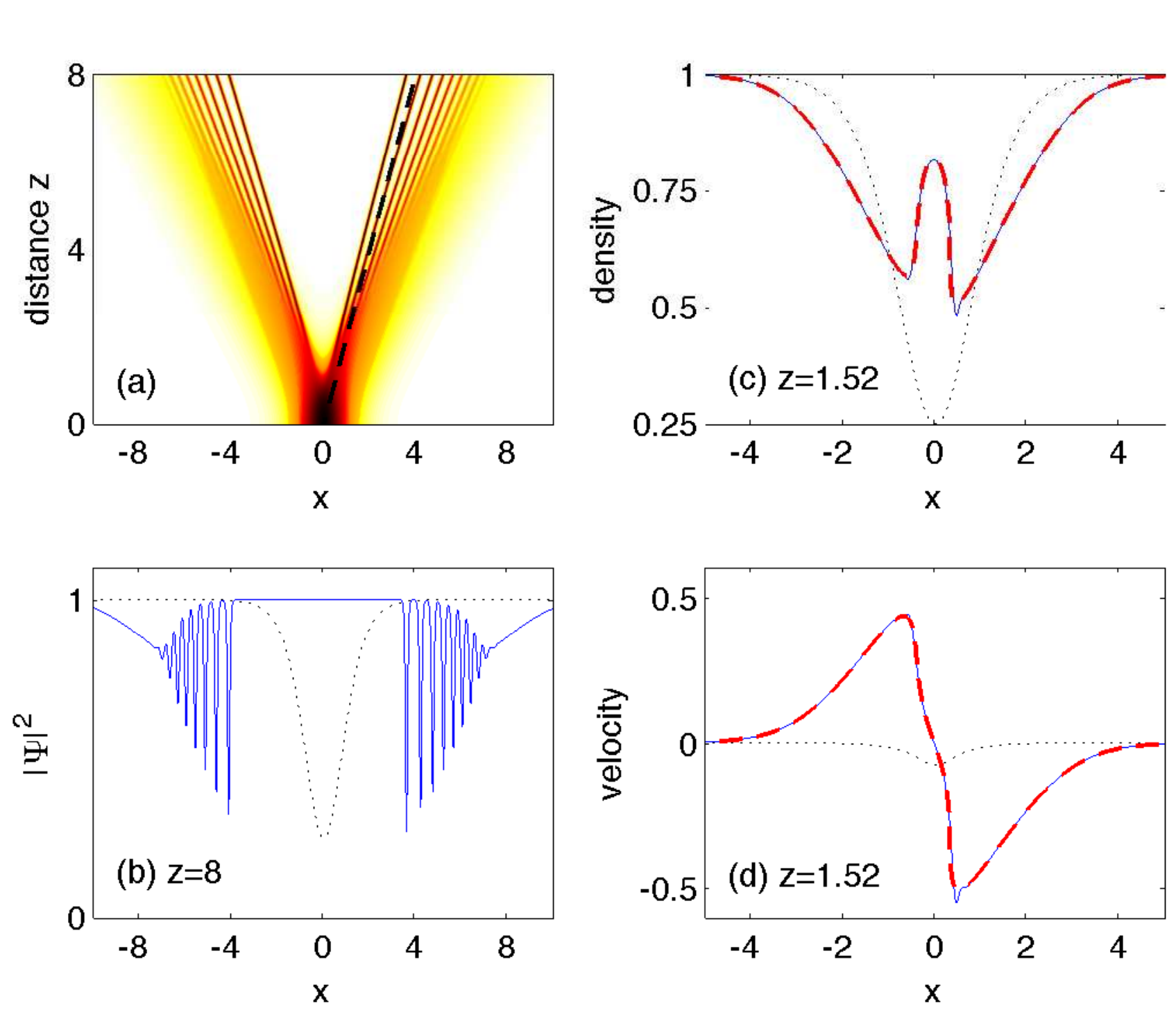}
%\includegraphics[width=8cm]{figpra/loc1D_v05}
% from data ...
\hfill
\caption{(Color online) DSW dynamics in a local medium ($\sigma=0$) arising from a gray soliton shape $u_0=\sqrt{1-v^2} \tanh(\sqrt{1-v^2} x) + i v$, with $v=0.5$. (a-b) Results from numerical integration of Eqs.~(\ref{nls3})-(\ref{nls4}) in 1+1D, with $\varepsilon=0.05$: (a) the twin shock fan in the $x-z$ plane.  The dashed line indicates the natural velocity of the input in the limit $\varepsilon=1$ for which the input itself is a one-soliton; (b) Output intensity in the soliton trains (solid line) compared with the input intensity (dotted line). (c-d) Comparison of the results from Eqs.~(\ref{nls3})-(\ref{nls4}) (solid line; blue online) with those obtained in the hydrodynamic limit from Eqs.~(\ref{wkb1})-(\ref{wkb2}) (thick dashed line; red online): (c) equivalent fluid density $h=|\psi|^2$ vs. $x$; (d) equivalent fluid velocity $u=\varphi_x$ vs x.}
\label{DSWloc1D_v05} 
\end{figure}
%-----------------------
Importantly, however, in a nonlocal medium, the stabilizing effect of the nonlocality on dark solitons allows us to envisage the stable formation of a regular DSW. This is demonstrated in Fig.~\ref{DSWnloc}, which is obtained by integrating  Eqs.~(\ref{nls3})-(\ref{nls4}) with $\sigma=1$, $\varepsilon=0.05$,  and input $U(x)$ corresponding to the exact nonlocal soliton shape
(quite similar results, not shown for brevity, are obtained with hyperbolic tangent input). 
First, as shown in Fig.~\ref{DSWnloc}(a), the wave-breaking scenario and the shock fan look qualitatively similar to that of the local case without noise, except for an increased shock distance $z_s$ (already pointed out in Ref.~\cite{Conti09}), and an adiabatic change of the width of solitons in the train. In this case, the details of the 1+1D evolutions could be described by means of Whitham modulation theory, whose application does not require integrability of the governing equations. While this is clearly beyond the scope of this work, our goal here is to emphasize that the output transverse pattern of intensity exhibits no transverse break up [see Fig.~\ref{DSWnloc}(b)] in spite of the noisy input. Conversely, the DSW develop in such a way that soliton filaments in the fan remain invariant stripes along $y$. Once seen at a fixed elevation $y$ the DSW looks as a soliton train which maintain close similarity to the train that one obtains when the propagation is strictly 1+1D \cite{Conti09}. In summary, the comparison of Fig.~\ref{DSWloc} and  Fig.~\ref{DSWnloc} allows us to conclude that the nonlocality can play a crucial role, 
enabling effective suppression of snake instabilities and observation of effective 1+1D DSW in a 1+2D setting. 
The effectiveness of such suppression mechanism is witnessed also by the fact that we observe the same scenario if we increase the initial noise level substantially (i.e., one order of magnitude) with respect to the local case reported in Fig. \ref{DSWloc}(d-f).
Such suppression is effective also for $\sigma<1$ (results not shown) and becomes more and more pronounced as $\sigma$ increases.
%----------------------- FIGURE 8 ------------------------------
 \begin{figure}[h]
 \includegraphics[width=5cm]{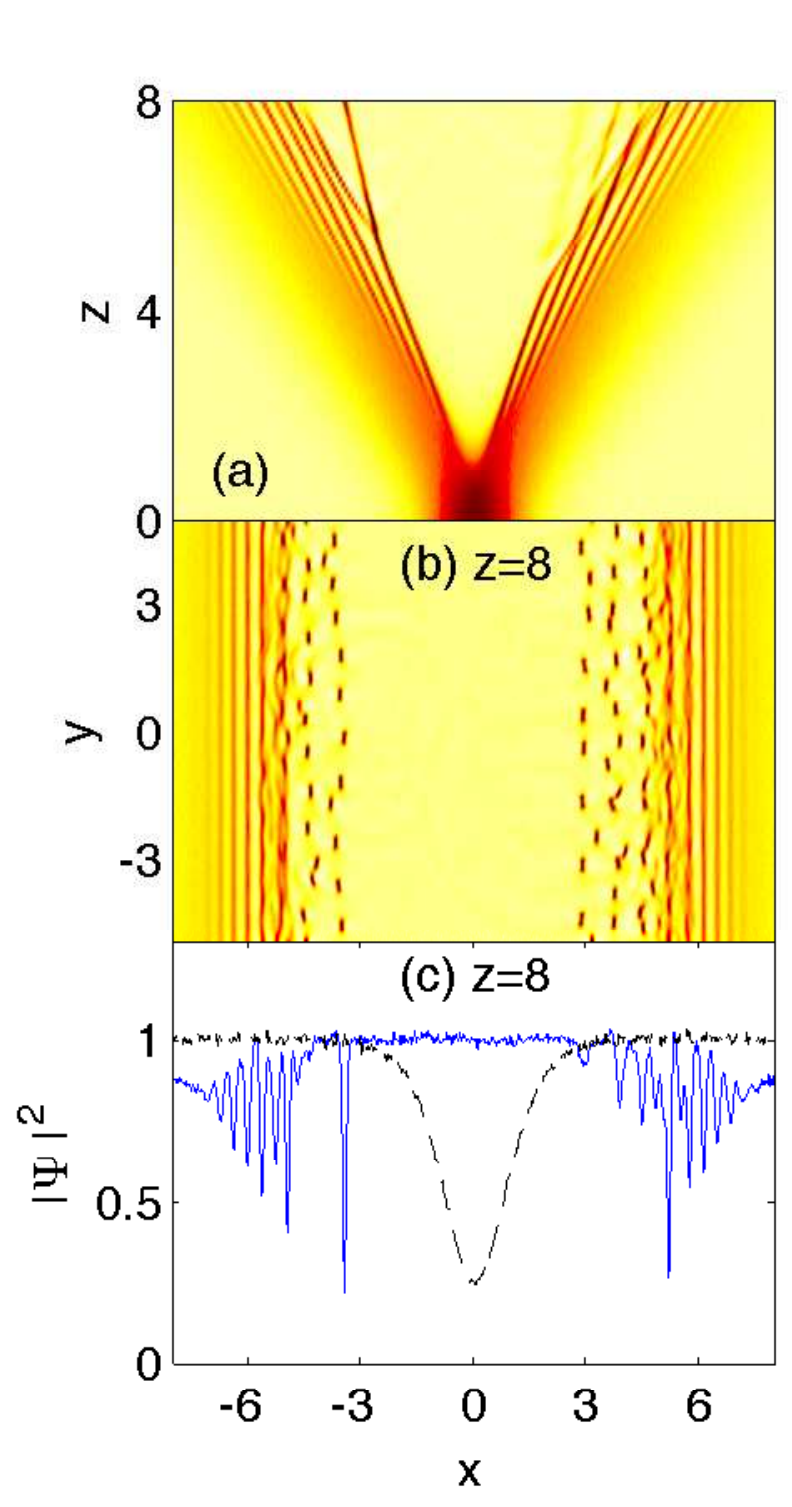} % Ste  from data 
\hfill
\caption{(Color online) As in Fig.~\ref{DSWloc1D_v05} in 1+2D, showing the onset of the instability growing from noise.
(a) Shock fan at $y=0$; (b) Output transverse plane; (c) Output intensity profile at $y=0$ (solid line)
compared with the input (dashed line).}
\label{DSWloc_v05} 
\end{figure}
%-----------------------

%----------------------- FIGURE 11 ------------------------------
 \begin{figure}[h]
 \includegraphics[width=5cm]{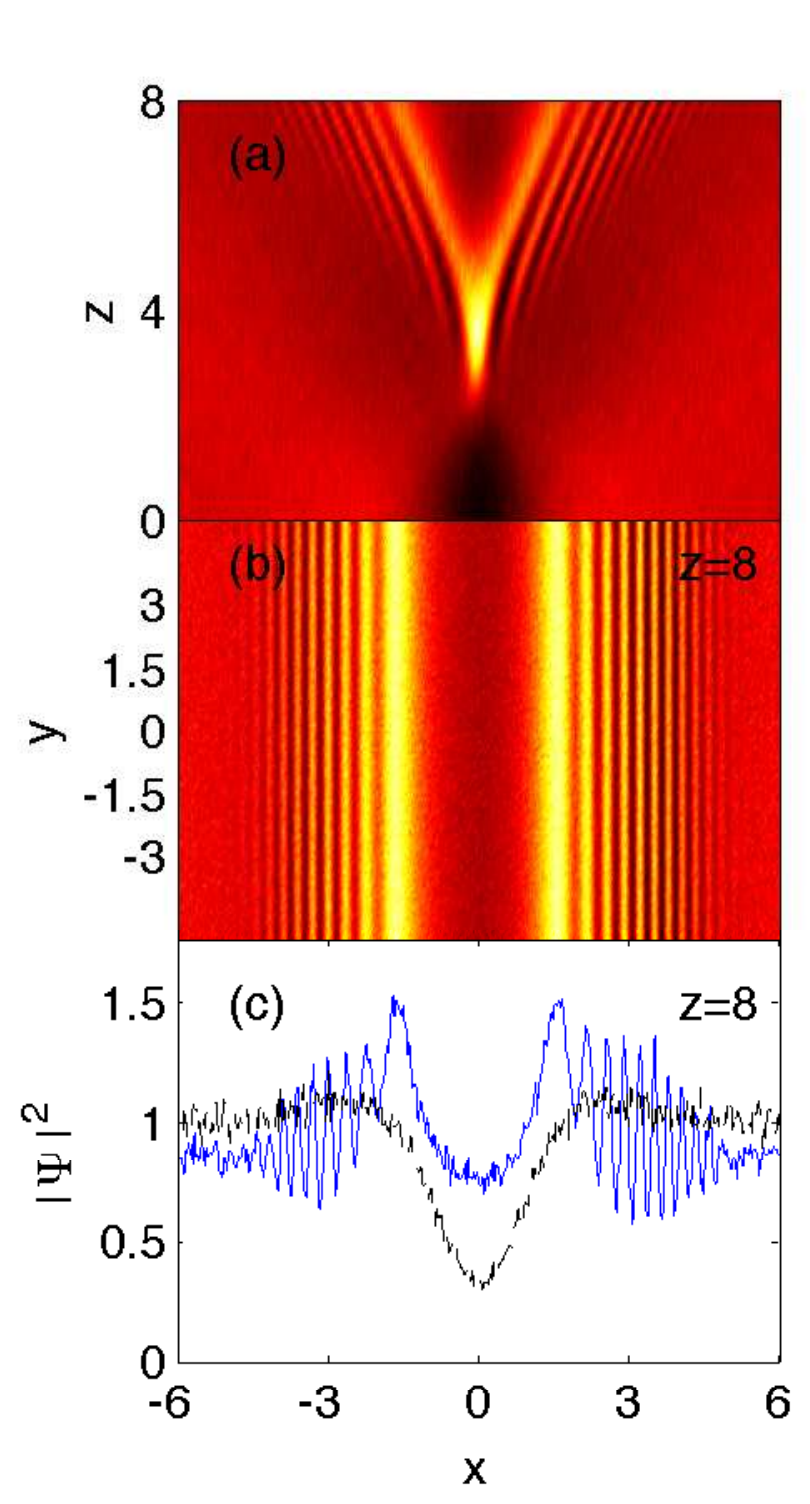}
%\includegraphics[width=5cm]{figpra/nloc_v05_sig1_modph0p02}
% from data nloc_v05_sig1_noise0p01.dat
%\includegraphics[width=8cm]{nloc_v04}	% Andrea
%\includegraphics[width=7cm]{figpra/nloc_v05_sig1_noise0p01}
\caption{(Color online) Stabilized DSW in the nonlocal case with $\sigma=1$.
Results from numerical integration of Eqs.~(\ref{nls3})-(\ref{nls4}) with $\varepsilon=0.05$
and  input shape characteristic of the nonlocal soliton $U(x)$ corresponding to $v=0.5$.}
\label{DSWnloc_v05} 
\end{figure}
%-----------------------
Finally, we have extended the numerical calculations in order to investigate the case of a dark input with non-zero minimum intensity.
First, we point out that, in this case, the wave-breaking scenario that leads to the appearance of the DSW changes qualitatively, as illustrated in Fig.~\ref{DSWloc1D_v05} in the local case.  We employ an input waveform corresponding to  a gray soliton $U(x)=\sqrt{1-v^2} \tanh(\sqrt{1-v^2} x) + i v$ of the integrable NLS equation. In this case, as shown in Fig.~\ref{DSWloc1D_v05}(a), a central black still soliton no longer appears. Conversely two distinct fans develop, which correspond to two non-symmetric soliton trains which split up on propagation, having net velocities of opposite sign [see Fig.~\ref{DSWloc1D_v05}(b)]. The two fans originates from regularization of two distinct shocks which develop trough gradient catastrophys at points $x=x_s$, located asymmetrically on the right ($x_s>0$) and left ($x_s<0$) of the input intensity dip (in $x=0$), respectively. These shocks develop at slightly different distances $z_s$, as one can predict from numerical solutions of the hydrodynamic limit [Eqs.~(\ref{wkb1})-(\ref{wkb2})]
\footnote{We integrate Eqs.~(\ref{wkb1})-(\ref{wkb2}) by means of up-wind finite difference schemes which give reliable results up to the gradient catastrophe point, beyond which the shock can be described appropriately only by means of shock-following algorithms.}. 
Snapshots of the field at distance $z_s=1.52$ corresponding to the formation of the right shock at $x_s>0$ are shown in Fig.~\ref{DSWloc1D_v05}(c-d), while the left shock  occurs at a slightly longer distance. In particular Fig.~\ref{DSWloc1D_v05}(c-d) compare the density $h=|\psi|^2$ and velocity $u=\varphi_x$ of the equivalent fluid, as obtained by integration of reduced hyperbolic [Eqs.~(\ref{wkb1})-(\ref{wkb2})] and the full model [Eqs.~(\ref{nls3})-(\ref{nls4})], respectively. As shown, the agreement is remarkably good up to the distance considered, beyond which the dynamics ruled by the full model deviates qualitatively starting to exhibit the soliton train features illustrated in Fig.~\ref{DSWloc1D_v05}(a-b). It is worth pointing out that the asymmetry between the right and left shocks is due to the soliton phase (implicitly considered in the initial condition $U(x)$), which is responsible for the initial profile of fluid velocity (i.e., the derivative of the phase with respect to $x$) reported in Fig.~\ref{DSWloc1D_v05}(d) as a dotted line. If such phase is suppressed, the two shocks turn out to be symmetrically located in $x$ around the origin, and develop at the same distance. 

As shown in Fig.~\ref{DSWloc_v05}, we find that the 1+1D scenario illustrated before changes once the propagation occurs in 1+2D, similarly to the case of black soliton input.
In fact, in 1+2D, the two DSW fans exhibit transverse break-up [Fig.~\ref{DSWloc_v05}(b)] due to the onset of the instability from noise. As a result, once the dynamics is seen on a plane $y=constant$, the solitons in the fans show erratic changes of velocity, as can be seen in Fig.~\ref{DSWloc_v05}(a). 
Note that the instability develops from the most unstable soliton stripes (i.e., the inner ones) in the two fans. When seen at given distance $z$ and transverse location $y$, the soliton trains appear irregular in $x$, as displayed in Fig.~ \ref{DSWloc_v05}(c). Moreover the pattern shown in Fig.~ \ref{DSWloc_v05}(c) changes from plane to plane $y=constant$. 
  
Importantly, however, when nonlocality becomes effective the twin shock fan is stabilized. A typical example, corresponding to the internal parameter value $v=0.5$, already employed in the local case [Fig.~\ref{DSWloc1D_v05} and Fig.~\ref{DSWloc_v05}], is illustrated in Fig.~\ref{DSWnloc_v05}.  In this case the input $U(x)$ corresponds to the exact soliton shape obtained with $\varepsilon=1$, for the external parameter $\sigma=1$ and $v=0.5$. In spite of a noise level much higher than the corresponding one in the local case, the DSW does not break up, showing that the suppression of the instability is effective also in this case. Similar results are obtained for different values of soliton velocity $v$, allowing us to conclude that the suppression of the instability due to nonlocal effect is a general phenomenon that permits to observe true 1+1D (stripe) DSW dynamics also in 1+2D under different initial conditions.

\section{Conclusions}
In summary we have shown that nonlocality plays a beneficial role in the propagation of dark stripes, 
allowing for a substantial suppression of transverse (snake) instabilities for both a single soliton and 
dispersive shock waves characterized by the emergence of multiple dark soliton stripes.
The residual weak instability leads, over long distances, to decay into vortex pairs, 
as numerical simulations of the nonlocal model seem to indicate. 

We point out that the nonlocal mechanism of suppression investigated in this paper, 
could cooperate, under suitable conditions, 
with other mechanisms previously investigated \cite{suppression}.
Moreover, stabilization due to nonlocality of dark stripes is expected also with focusing nonlinearities, 
when the signs of second derivatives in the transverse plane are opposite (hyperbolic case).

\section{Acknowledgement}
The authors acknowledge illuminating discussions with C. Conti.
Andrea Fratalocchi's research (www.solarpaintproject.org) is supported by Award No. KUK-F1-024-21 (2009/2012), 
made by King Abdullah University of Science and Technology (KAUST). 

%\begin{thebibliography}

%\end{multicols}
\end{document}